\documentclass[aps,preprint,superscriptaddress,amsmath,amssymb,pre,floatfix]{revtex4-1}

\usepackage{hyphenat}
\usepackage{hhline}
\usepackage{upgreek}
\usepackage{graphicx}
\usepackage{soul,color}
    
\newcommand{\rmlabels}[3]{\ensuremath{#1^{\mathrm{#2}}_{\mathrm{#3}}}}

\newcommand{\td}{\ensuremath{T_{\mathrm{d}}}}
\newcommand{\tr}{\ensuremath{T_{\mathrm{rep}}}}
\newcommand{\ti}{\ensuremath{T_{\mathrm{int}}}}
\newcommand{\tpeak}{\ensuremath{T_{\mathrm{PtP}}}}

\def\longrightharpoonup{\relbar\joinrel\rightharpoonup}
\def\longleftharpoondown{\leftharpoondown\joinrel\relbar}

\def\longrightleftharpoons{
  \mathop{
    \vcenter{
      \hbox{
      \ooalign{
        \raise1pt\hbox{$\longrightharpoonup\joinrel$}\crcr
	  \lower1pt\hbox{$\longleftharpoondown\joinrel$}
	  }
      }
    }
  }
}

\newcommand{\tref}[1]{Table~\ref{table:#1}}

\newcommand{\tlabel}[1]{\label{table:#1}}

\newcommand{\fref}[1]{Fig.~\ref{fig:#1}}

\newcommand{\flabel}[1]{\label{fig:#1}}

\newcommand{\eref}[1]{Eq.~\ref{eqn:#1}}
\newcommand{\erefs}[1]{Eqs.~\ref{eqn:#1}}

\newcommand{\elabel}[1]{\label{eqn:#1}}

\begin{document}

\title{Robustness of synthetic oscillators in growing and dividing cells}

\author{Joris \surname{Paijmans}}
\affiliation{AMOLF, Science Park 104, 1098 XG Amsterdam, The Netherlands}
\author{David K. \surname{Lubensky}}
\affiliation{Department of Physics, University of Michigan, Ann Arbor, MI 48109-1040, USA}
\author{Pieter Rein \surname{ten Wolde}}
\affiliation{AMOLF, Science Park 104, 1098 XG Amsterdam, The Netherlands}

\begin{abstract}
  Synthetic biology sets out to implement new functions in cells, and
  to develop a deeper understanding of biological design principles.
  In 2000, Elowitz and Leibler showed that by rational design of the
  reaction network, and using existing biological components, they
  could create a network that exhibits periodic gene expression,
  dubbed the repressilator (Elowitz and Leibler, Nature, 2000). More
  recently, Stricker et al. presented another synthetic oscillator,
  called the dual-feedback oscillator (Stricker et al., 2008), which
  is more stable. How the stability of these oscillators is affected
  by the intrinsic noise of the interactions between the components
  and the stochastic expression of their genes, has been studied in
  considerable detail.  However, as all biological oscillators reside
  in growing and dividing cells, an important question is how these
  oscillators are perturbed by the cell cycle.  In previous work we
  showed that the periodic doubling of the gene copy numbers due
  to DNA replication can couple not only natural, circadian
  oscillators to the cell cycle (Paijmans et al., PNAS, \textbf{113}, 4063, (2016)), 
  but also these synthetic oscillators.  Here we expand this study. 
  We find that the strength of the locking between oscillators depends 
  not only on the positions of the genes on the chromosome,
  but also on the noise in the timing of gene replication: noise tends
  to weaken the coupling. Yet, even in the limit of high levels of noise
  in the replication times of the genes, both synthetic oscillators show clear
  signatures of locking to the cell cycle.  This work enhances our
  understanding of the design of robust biological oscillators inside
  growing and diving cells.
\end{abstract}


\maketitle

\section{Introduction}
Synthetic biology strives to implement new functions in living cells,
and to develop a deeper understanding of biological design principles,
using a modular rational design of biochemical reaction networks
\cite{Elowitz2000,Collins2000,Becskei2000}. As synthetic biology
becomes more mature, the goal is to design robust, stable and tunable networks
\cite{Andrianantoandro2006,Tigges2009,Nandagopal2011,Sowa2015}, which
are resilient to the effects of intrinsic noise and stochastic gene expression
\cite{Elowitz2002a,Elowitz2002,Rosenfeld2005,Cai2006,Chabot2007}.
In oscillators, enhanced robustness has been achieved via
the design of the reaction network at the single cell level
\cite{Stricker2008,Tsai2008,Novak2008,Mather2009,Woods2016,Potvin-Trottier2016}, 
and by connecting multiple cells through quorum sensing
\cite{Garcia-Ojalvo2004,Mondragon-Palomino2011,Prindle2014}.  
These analyses, however, have generally ignored a potentially major source of perturbation to synthetic oscillators:
The periodic gene replication and cell division that occur in any growing cell \cite{Feillet2014, Bieler2014}.
Cell division introduces noise due to the binomial partitioning of the proteins \cite{Kang2008, Gonze2013}. 
Moreover, we recently showed that circadian oscillators can lock to the cell cycle via the
periodic discrete gene duplication events arising from DNA replication during the cell cycle \cite{Paijmans2016}.
Here we study in detail how two synthetic oscillators are affected by the cell cycle,
and especially by these discrete replication events.
  
The mechanism by which cellular oscillators can couple to the cell cycle is generic and pertains 
to any biochemical oscillator in growing and dividing cells. 
Since the genes need to be replicated during the cell cycle,
and because the transcription rate is often proportional to the gene copy number in a cell \cite{Rosenfeld2005, Walker2016},
the cell cycle can cause a periodic doubling in the transcription rate of the clock related genes.
While the mechanism of coupling is generic, it is best understood in the context of an oscillator consisting of one
clock protein, which is a transcription factor that negatively autoregulates the expression of its own gene \cite{Paijmans2016}.
The periodic doubling of the gene copy number due to DNA replication leads to a
periodic doubling of the gene density. This means that the synthesis rate 
of the clock protein depends on the phase of the clock with respect to
that of the cell cycle: if the gene is expressed
when its gene density is maximal, then the amplitude of the protein
concentration will be maximal too. This increases the
amplitude of the oscillation, and since the subsequent decay of the
protein concentration does not depend on the gene density, the
rise in amplitude will increase the period of the
oscillation. The period of the oscillation thus depends on the phase
of the oscillator with respect to that of the cell cycle, and this
allows, as for any nonlinear oscillator, 
the cell cycle to strongly influence the synthetic oscillator \cite{Pikovsky2003}.


The two synthetic oscillators that we study are the repressilator,
developed by Elowitz and Leibler \cite{Elowitz2000}, and the
dual-feedback oscillator, developed by Stricker and coworkers \cite{Stricker2008}. 
Both oscillators have been reconstructed in {\it E. coli}. 
In our previous work, we showed by mathematical modeling that both oscillators 
can lock to the cell cycle \cite{Paijmans2016}. 
Also the authors of \cite{Dies2015} found, independently,  
by combining modeling with experiments, that the
dual-feedback oscillator can be entrained by the cell cycle.  Here
we study how the coupling strength depends on the noise in gene
replication, and, following earlier work \cite{Paijmans2016}, on
the positions of the genes on the DNA.  We modify the original
computational models of the repressilator and dual-feedback
oscillator, to include the periodic doubling of the mRNA production
rate with the cell cycle.  We consider the scenario that the
synthetic oscillators are incorporated into the chromosome, although
we will also discuss the fact that in the experiments the
oscillators are implemented on plasmids present at high copy number
\cite{Elowitz2000,Stricker2008}.  Under typical slow growth
conditions, {\it E. coli} has one chromosome at the beginning of the
cell cycle, in which case the gene copy number goes from 1 to 2 over
the course of the cell cycle.  At high growth rates, corresponding
to cell division times shorter than the replication time of the DNA
(on the order of 40 minutes), the chromosome can have multiple
replication forks, which means that the gene copy number can be
larger.  Here, we only consider the regime that the cell division
time is on the order of the DNA replication time or longer, such
that the gene copy number rises from $N=1$ at the beginning of the
cell cycle to $2N=2$ at the end.  To quantify the sensitivity of the
network to the cell cycle, we investigate the effect on the
peak-to-peak time in the protein concentrations related to the
oscillator, for different periods of the cell cycle.

Unlike the Kai circadian clock, these two genetic oscillators
comprise more than one operon that shows significant time variation in its expression. 
This introduces new important timescales to the problem: If the genes pertaining to the oscillator
are placed at a distance on the chromosome, there is a time delay
between when they are replicated.  The synthetic oscillators studied
here have an intrinsic period that is on the order of hours
\cite{Elowitz2000, Stricker2008}, which is similar to the timescale of
DNA replication, which takes at least 40 minutes.  Consequently, the
time delay can, depending on the reaction network, have a strong
effect on the period of the oscillations.

Both synthetic oscillators can lock to the cell cycle for a wide range
of cell division times, but, as we reported in our earlier work
\cite{Paijmans2016}, the effect critically depends on the positioning
of the genes on the chromosome: Where the repressilator almost shows
no locking when the genes are placed adjacently, the dual-feedback
oscillator, to the contrary, experiences the strongest effect in this
case, and locking decreases as the genes are placed further apart.

The pronounced effect of varying the delay between replication of different genes 
suggests that synthetic oscillators should also be sensitive to stochastic variation in replication times.
Our major goal here is thus to understand how such variation contributes to noise in the period of cellular oscillators.
The noise in the replication time is the result of two stochastic processes: 
The timing of initiation of DNA replication and the progression of DNA
replication.  Stochasticity in the initiation of replication has the
same effect on all the genes on the chromosome; a fluctuation in the
initiation time propagates to the replication times of all the genes,
leaving the interval between the gene replication times unchanged.  In
contrast, stochasticity in replication progression introduces temporal
fluctuations in the time between the replication of different genes.


Our simulation results show that, for physiological levels
for the noise in the gene replication times, 
the effect of gene replications on the period of the oscillations 
are strongly attenuated.
However, clear signatures of the cell cycle are observable,
especially around the 1:1 locking region.
We then address the question which noise source has 
the strongest effect on attenuating the effects of the cell cycle:
Initiation or progression of DNA replication.
To find out, we study the effects of the cell cycle in two different scenarios:
Either there is noise in the initiation of replication,
such that the timing between replicating different genes is fixed,
or the noise is limited to the progression of replication
such that the initiation time is fixed and the timing between genes is stochastic.
Our results reveal that noise in the initiation of DNA replication 
reduces the effect of locking much more than noise in DNA replication progression.  
This is because at biologically relevant noise levels, 
the standard deviation in the initiation of DNA replication is much 
larger than that in the progression of replication.
Nevertheless, even with high levels of noise in the initiation of DNA replication, 
the effects of locking are still clearly present for cell division times around
the oscillator's period.  Our results thus predict that synthetic
oscillators will be perturbed by the cell cycle in growing and
dividing cells, when the oscillators are implemented on the chromosome.

Below, we first give an overview of the models for the repressilator,
the dual-feedback oscillator and the models for the cell cycle. 
First we give a description of a completely deterministic cell cycle, 
and then introduce stochasticity in the model by making
the time DNA replication is initiated and the time it takes to replicate the DNA stochastic variables.
To determine how strong the oscillators are coupled to the cell cycle,
we study how the period of the oscillators scales with the cell division time.


\section{Theory}
To study the effect of the cell cycle on synthetic oscillators, 
we will use the ODE models of the repressilator \cite{Elowitz2000} 
and dual-feedback oscillator \cite{Stricker2008}, 
as described in these papers. As we argue in
more detail in \cite{Paijmans2016}, the key quantity connecting the
cell cycle and the oscillator is the gene density, $G(t)$,
i.e. the gene copy number per unit cell volume. 
Because the protein production rate is proportional to the gene copy number, 
discrete gene replication events cause sudden doubling of the production rate
(at least in prokaryotes \cite{Walker2016,Barziv2016}). 
We include the effects of the discrete gene replication events 
by making the mRNA productions rate due to transcription of each gene $i$
proportional to the gene density $G_{i}(t)=g_{i}(t)/V(t)$ \cite{Paijmans2016}. 
Here $g_{i}$ is the gene copy number of gene $i$ which switches from 1 to 2 during the
cell cycle, and $V(t)$ is the cell volume which exponentially doubles
in size during a cell division time \td.

\subsection{Repressilator} 
The repressilator consists of three genes, which sequentially repress
each other's expression.  As schematically shown in
\fref{Model_Cartoons}A, the first gene represses the expression of the
second, which represses the third gene, which in turn represses the
expression of the first again \cite{Elowitz2000}.  To take into
account gene replication, the expression of mRNA is proportional to
the gene density $G_i(t)$
\begin{eqnarray}
 \frac{dm_{i}(t)}{dt} & = & - m_{i}(t) + \frac{G_i(t)}{\bar{G_i}} \frac{\alpha}{1+(p(t)_{i-1})^n} + \alpha_0 \\
 \frac{dp_{i}(t)}{dt} & = & -\mu_p p_{i}(t) + \gamma m_{i}(t) \nonumber.
\end{eqnarray}
Here, $m_{i}$ and $p_i$ are the concentrations of mRNA and proteins
($i\in\{1,2,3\}$), respectively, both rescaled with the constant of half-maximum
repression $K_{M}$.  The transcription rate is assumed to be
proportional to the instantaneous gene density $G_i(t)$; importantly,
the gene density can differ between the three genes when they are
positioned differently on the chromosome, see \fref{Model_Cartoons},
panels B and C.  $\bar{G}_i$ is the time-averaged gene density, which
depends on the phase of the cell cycle at which the gene is
duplicated.  The mRNA expression has a basal rate $\alpha_0$ and an
enhanced rate $\alpha$, which is repressed by protein $p_{i-1}$, 
where $i-1$ is mod 3, with a Hill coefficient $n$; here, following the
original paper \cite{Elowitz2000}, time is rescaled in units of the
mRNA lifetime and protein concentrations are in units of the
concentration necessary for half-maximal repression.  In the second
equation, $\mu_p$ is the protein decay rate over the mRNA decay rate
and $\gamma$ is the translation efficiency, {\it i.e.} the average
number of proteins produced per mRNA molecule.  We used the parameters
given in \tref{CHS_Parameters}.

\begin{figure}[ht]
\center
\includegraphics[scale=0.95]{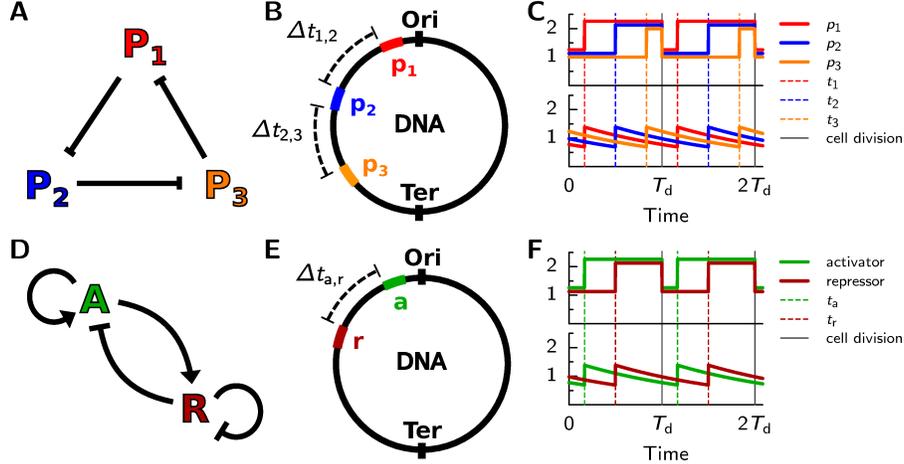}
\caption{\flabel{Model_Cartoons} Models for the synthetic oscillators and cell cycle.
(A) Network architecture of the Repressilator \cite{Elowitz2000}: 
$P_1$ represses the production of $P_2$, $P_2$ represses $P_3$ and $P_3$ represses $P_1$ again. 
(B) Illustration of circular chromosome, with the origin (Ori) and termination (Ter) of replication.
When the three genes $p_1$,$p_2$ and $p_3$ are placed at a distance on the chromosome,
there are temporal delays $\Delta\rmlabels{t}{}{1,2}$ and $\Delta\rmlabels{t}{}{2,3}$,
between when the genes are replicated.
(C) Gene copy numbers (top) and gene densities (bottom) 
of the genes $p_1$ (red), $p_2$ (blue) and $p_3$ (orange), respectively. 
They are replicated at times $t_1$, $t_2$ and $t_3$, respectively, as indicated by the dashed lines.
The black vertical lines indicate cell divisions.
For the gene copy number, lines are shifted vertically for clarity.
(Bottom) Gene densities for each gene, normalized by their average.
(D) Network architecture of the dual-feedback oscillator \cite{Stricker2008}:
The activator (A) auto-activates its production and enhances the production of the repressor (R).
The repressor auto-represses its production, and suppresses the
production of the activator. 
(E) Schematic of the circular chromosome.
The genes for the activator (a) and repressor (r) are placed at different positions on the DNA, 
such that there is a temporal delay, $\Delta\rmlabels{t}{}{a,r}$, between there respective replication times.
(F) Gene copy numbers (top) and gene densities (bottom) of the genes a (green) and r (red), respectively.
Genes a and r are replicated at times, $t_\mathrm{a}$ and $t_{\mathrm{r}}$, respectively, 
indicated by the dashed vertical lines.
The black vertical lines indicate cell divisions.
For the gene copy number, lines are shifted vertically for clarity.
}
\end{figure}

\begin{table}
\begin{center}
    \begin{tabular}{ c r l } 
     {\bf Parameter} & {\bf Value} & {\bf Definition and motivation} \\ \hline \hline
     \rmlabels{\alpha}{}{init} & 0.2 & Fraction of \td{} when replication starts \cite{Wallden2016}. \\
     \tr & 40 min & Mean DNA replication time in {\it E. coli}. \\
     $\Delta\rmlabels{t}{}{1,2}$/\tr & 0,$\frac{1}{14}$,$\frac{1}{5}$,$\frac{1}{2}$ & Time between gene replications (repressilator) \\     
     $\Delta\rmlabels{t}{}{a,r}$/\tr & 0,$\frac{1}{8}$,$\frac{1}{2}$,1 & Time between gene replications (dual-feedback) \\
     \rmlabels{\sigma}{}{rep} & 0.35\,\tr{} & SD in DNA replication progression \cite{Adiciptaningrum2015}. \\
     \rmlabels{\sigma}{}{init} & 0.20\,\td{} & SD in initiation of DNA replication  \cite{Adiciptaningrum2015}. \\
     arab & 0.70\% & Arabinose level in dual-feedback oscillator. \\
     $[$IPTG$]$ & 2nM & IPTG concentration in dual-feedback oscillator. \\ \hline
    \end{tabular}
\end{center}
\caption{\tlabel{CHS_Parameters} Parameters used in the models. 
For the repressilator, we used the parameters given in Box 1 in \cite{Elowitz2000}.
For the dual-feedback oscillator we used the parameters given in the SI of \cite{Stricker2008}.
SD stands for standard deviation.}
\end{table}

\subsection{Dual-feedback oscillator}
The dual-feedback oscillator, schematically shown in
\fref{Model_Cartoons}D, consists of two genes, one coding for an
activator and one for a repressor \cite{Stricker2008}.  The activator
enhances the expression of both genes, while the repressor represses
the expression of both genes.  Since the genes have identical
promoters, the temporal expression of the two proteins is similar.
The model we employ is presented in the SI of \cite{Stricker2008}, but
to take into account the periodic variations in the gene density, we
have modified the equations describing the transcription of mRNA of
the activator and repressor
\begin{eqnarray}
 P_{0,0}^{a/r} \xrightarrow{\frac{b_{a/r}}{\bar{G}_{a/r}}G_{a/r}(t)} P_{0,0}^{a/r} + m_{a/r} \\
 P_{1,0}^{a/r} \xrightarrow{\frac{\alpha b_{a/r}}{\bar{G}_{a/r}}  G_{a/r}(t)} P_{1,0}^{a/r} + m_{a/r}. \nonumber
\end{eqnarray}
Here $P_{m,n}^{a/r}$ denotes the promoter of the (a)ctivator/(r)epressor gene, 
with $m=0,1$ activator protein and $n=0,1$ repressor protein bound to it, respectively. 
The mRNA $m_{a/r}$ of the activator ($a$) and repressor ($r$) is transcribed with a rate $(\alpha) b_{a/r} G(t)$, 
which depends on the state of the promoter and on the gene density $G_{a/r}(t)$. 
Parameters are given in \tref{CHS_Parameters}. 
Genes can be placed at a distance from each other on the chromosome,
as shown in \fref{Model_Cartoons}E and F,
which introduces a delay between when they are replicated.
The intrinsic period of this oscillator without the driving by the gene density is $\sim 40$ minutes 
and we want to study the behavior of the oscillator in a wide window of cell division times 
around this period.
Because in our model of the cell cycle \td{} always needs to be longer than the 
DNA replication time of 40 minutes,
it is convenient to study the dual-feedback oscillator with an intrinsic period that is longer than the current 40 minutes.
To obtain a longer clock period, we use the experimental observation in \cite{Stricker2008} 
that the clock period scales with temperature via the Arrhenius law. 
To this end, we scale all rate constants, $k_i$, in the model, using 
\begin{equation}
  k_i = k_{\mathrm{ref}}\,\mathrm{exp}(-\Theta_{\mathrm{cc}} [1/T-1/T_{\mathrm{ref}}]),
\end{equation}
where $k_{\mathrm{ref}}$ is the rate constant at the reference temperature
$T_{\mathrm{ref}}$ of 310K and $\Theta_{\mathrm{cc}}\approx8300$K is a constant.
We will evaluate the model at a temperature of 303K where the clock has an intrinsic period of about 73 minutes.

\subsection{Cell cycle model}
The time at which a gene is replicated depends
on the timing of two major events, which divide the cell cycle into three distinct intervals: 
The time between the start of the cell cycle and initiation of DNA replication,
the replication time of the chromosome and, after this has finished,
the time until cell division.  As we argued in \cite{Paijmans2016},
cell division has a smaller effect on the oscillator as compared to
gene replication, as both the cell volume and the gene copy number
divide by two at cell division, leaving the important gene density
unchanged.  Therefore, in our model we assume there is no stochasticity in
the division time \td, which we keep fixed. Furthermore, we assume that
the {\it E. coli} cells grow slowly, such that the division time is
always longer than the DNA replication time.  In this case, there are
at most two origins of replication per cell, and we do not have to
take into account the effects of multiple replication forks
\cite{Cooper1968}.

Because it is still poorly understood how the cell coordinates the
replication and division cycles, in this work we employ a simple
model for the cell cycle.  Evidence emerges that initiation of
chromosome replication is triggered at a fixed density of the origin
of replication (Ori), $\rmlabels{\hat{G}}{}{Ori}$, independent of
cell's division time \cite{Donachie1968,Wallden2016}. 
Given that the density of the Ori depends on the cell volume $V(t)$,
$\rmlabels{G}{}{Ori}=1/V(t)$, the time and precision of initiation of
DNA replication is set by the evolution of the cell volume and the
precision of the sensor for $\rmlabels{G}{}{Ori}(t)$.  
Because we consider the slow growth regime
where at the beginning of the cell cycle there is only one origin of
replication, and because we assume that the initial volume is
independent of the growth rate, it follows that the average time at which DNA
replication is initiated is at a fixed fraction
\rmlabels{\alpha}{}{init} of the division time \td,
$\Delta\rmlabels{t}{}{init}=\rmlabels{\alpha}{}{init}\td$, 
with a standard deviation \rmlabels{\sigma}{}{init}.
We choose, based on data presented in \cite{Wallden2016}, $\rmlabels{\alpha}{}{init}=0.2$. 
The time it takes to replicate the chromosome depends on the speed of the DNA polymerase, 
which in turn can depend on the cell's physiological state \cite{Adiciptaningrum2015}. 
For simplicity, we assume that the mean time to replicate the whole
chromosome is $\tr=40$ minutes with a standard deviation given by \rmlabels{\sigma}{}{rep},
both independent of the cell's division time.
In this work we consider two models for the timing of gene replications:
One where both the initiation and the progression of DNA replication
are deterministic, such that gene replications occur at the same phase
each cell cycle and one where we introduce noise in these two processes.
The effects of noise in the initiation and progression of DNA replication 
on the gene replication times is illustrated in \fref{CellCycle_Cartoons}.

\begin{figure}[ht]
\includegraphics[scale=1.0]{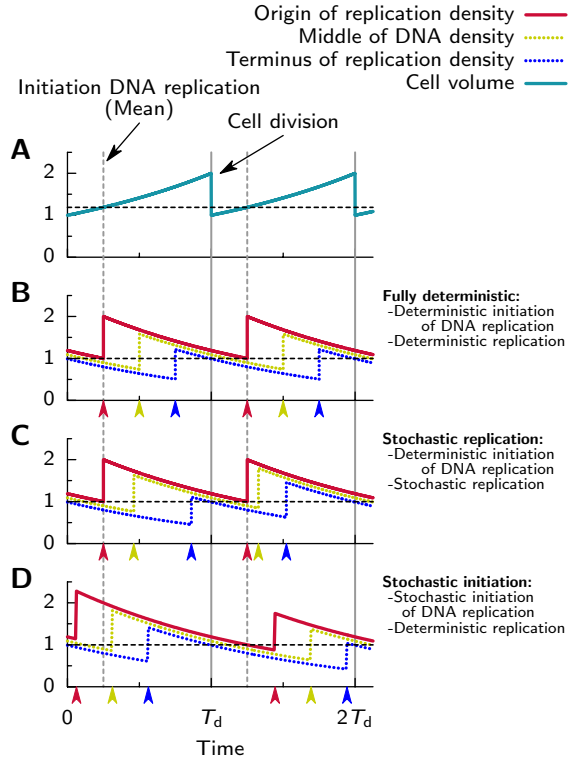}
\caption{\flabel{CellCycle_Cartoons} Models to determine the gene replication times.
(A) Time trace of cell volume, which is a deterministic function of time in all models, 
where cell division occurs with a period \td{}, indicated by the vertical solid gray lines.
The vertical dashed gray lines indicate the times at which DNA replication is initiated
when the timing of the initiation of replication is deterministic. 
The horizontal dashed lines show the volume (A) or the concentration of the origin of replication (B-D),
at which DNA replication, on average, initiates.
(B-D) Time traces of the density of the origin of replication of the chromosome (red solid line),
a gene precisely half-way the origin and terminus of replication (green dotted line) 
and the terminus of replication (blue dotted line).
Arrows below the x-axis indicate the replication times of these sites.
Note that the gene densities show no discontinuity at cell division.
All gene densities are normalized by the critical density for replication initiation.
(B) Fully deterministic model. Initiation of replication and the replication of the two genes occur at fixed times each cell cycle.
(C) When there is stochasticity in DNA replication progression, 
the timing between initiation of replication and the replication of genes further along the DNA becomes stochastic.
(D) When the initiation of replication is stochastic, but the replication rate is constant,
all replication events move in concert, and the time between initiation and replication of the genes is fixed.
}
\end{figure}

\subsubsection{Deterministic model}
The first model is completely deterministic. 
Indeed, when we assume the evolution of the cell volume, $V(t)$, 
to be deterministic and that DNA replication initiates exactly when 
$\rmlabels{G}{}{Ori}(t)=\rmlabels{\hat{G}}{}{Ori}$, 
then the evolution of $\rmlabels{G}{}{Ori}(t)$ becomes fully deterministic.
Clearly, since both the initiation and the progression of DNA replication are
deterministic, the respective genes are copied at the same times each
cell cycle (See \fref{CellCycle_Cartoons}A and B).  
Furthermore, in our model the first gene of the oscillator is next to the
origin of replication, such that the time this gene is replicated,
$t_1=\Delta\rmlabels{t}{}{init}=\rmlabels{\alpha}{}{init}\td$. 
Note that it is not important when exactly during the cell cycle the gene
is replicated, as it only changes the gene density by a prefactor,
which we compensate for by normalizing $G_{i}(t)$ by its mean $\bar{G}_{i}$. 
However, as we will see, the time between the replication
of the different genes is important. 
Genes can be placed apart on the DNA which introduces a time delay, $\Delta t_{i,j}$, 
between when the genes $i$ and $j$ are copied, respectively. 
The times during the cell cycle when the genes $p_1,p_2$ and $p_3$ are replicated, 
for the repressilator, and the activator and repressor genes for the
dual-feedback oscillator, are
\begin{align}
 \elabel{RepTimesDet}
 t_1 & = \Delta\rmlabels{t}{}{init}             & t_{\mathrm{a}} & = \Delta\rmlabels{t}{}{init} \\ \nonumber
 t_2 & = t_1 + \Delta t_{1,2}  & t_{\mathrm{r}} & = t_{\mathrm{a}} + \Delta t_{\mathrm{a,r}} \\ \nonumber
 t_3 & = t_2 + \Delta t_{2,3}  &                &  
\end{align}

%

\subsubsection{Stochastic model: Noise in the initiation and progression of DNA replication}
For the second model, we again assume that the evolution of the cell volume is deterministic,
but turn replication progression and replication initiation into stochastic processes.
Due to stochasticity in the progression of DNA replication,
the time interval between the gene replication events becomes stochastic,
as illustrated in \fref{CellCycle_Cartoons}C.
We assume the time it takes to replicate the full chromosome follows a Gaussian distribution with a mean \tr=40 min,
and standard deviation $\rmlabels{\sigma}{}{rep}$ that is proportional to replication time \tr.
When the standard deviation in the DNA replication time is the result of many independent stochastic steps,
the time between replicating genes $i$ and $j$, $\delta\tau_{i,j}$, 
which on average takes a time $\Delta t_{i,j}$,
will therefore also be Gaussian distributed with a standard deviation of $\sqrt{\Delta t_{i,j}/\td}\,\rmlabels{\sigma}{}{rep}$.

Stochasticity in the initiation of replication affects the replication times of all genes equally; 
indeed, the time between copying two different genes, $\Delta t_{i,j}$, is constant,
as is shown in \fref{CellCycle_Cartoons}D.
This stochasticity in the timing of the initiation can come from the sensing limit of measuring
$\rmlabels{G}{Ori}{}(t)$, or because of stochasticity in the evolution of the cell volume 
(which, however, we assume to progress deterministically in this scenario).
In our model, the time of initiation of DNA replication, $\delta\rmlabels{\tau}{}{init}$, 
is a stochastic variable drawn from a Gaussian probability distribution with 
a mean $\rmlabels{\alpha}{}{init}\td{}$ with a standard deviation \rmlabels{\sigma}{}{init}.
Assuming the standard deviation in measuring $\rmlabels{G}{}{Ori}(t)$, $\rmlabels{\sigma}{}{G_{\rm Ori}}$,
is small, the standard deviation in the initiation time is 
\begin{equation}
  \elabel{OriCuncertain}
  \rmlabels{\sigma}{}{init} = \left|\left.\frac{d(\Delta\rmlabels{t}{}{init})}{d\rmlabels{G}{}{Ori}}\right|_{\rmlabels{G}{}{Ori}=\rmlabels{\hat{G}}{}{Ori}}\right|\,\rmlabels{\sigma}{}{G_{\rm Ori}}.
\end{equation}
DNA replication is initiated when $\rmlabels{G}{}{Ori}=V_0^{-1}{\rm exp}(-{\rm ln}(2)/\td\,\Delta\rmlabels{t}{}{init})=\rmlabels{\hat{G}}{}{Ori}$,
where $V_0$ is the cell volume after cell division.
Solving this equation for the initiation time gives $\Delta\rmlabels{t}{}{init}=-\td{\rm ln}(\rmlabels{\hat{G}}{}{Ori}\,V_0)/{\rm ln}(2)$.
Then, from \eref{OriCuncertain} it follows that the standard deviation in the initiation time is
$\rmlabels{\sigma}{}{init}\sim \td\rmlabels{\sigma}{}{G_{\rm Ori}}$.
Therefore, in our model, the standard deviation in the initiation time is proportional to \td.

Assuming that the two stochastic processes are independent, 
the replication times of the genes for the repressilator and dual-feedback oscillator become, respectively
\begin{align}
 \elabel{RepTimesStochComb}
 t_1 & = \delta\rmlabels{\tau}{}{init} & t_{\mathrm{a}} & =  \delta\rmlabels{\tau}{}{init} \\ \nonumber
 t_2 & = t_1 + \delta\tau_{1,2}     & t_{\mathrm{r}} & =  t_{\mathrm{a}} + \delta\tau_{\mathrm{a,r}} \\ \nonumber
 t_3 & = t_2 + \delta\tau_{2,3}     &                &  
\end{align}
Because in our model, the division time is fixed each cell cycle,
we have to constrain the values of the replication times to lie within the finite interval $[0,\td]$.
First we choose $\delta\rmlabels{\tau}{}{init}$, and constrain it to lie within $[0,(\td-\tr)]$.
Then we draw a value for $\delta\tau_{1,2}$ and constrain it to lie within the interval 
that is symmetric around its mean value $\Delta\tau_{1,2}$, $[0,2\Delta t_{1,2}]$.
Similarly, we draw a value for $\delta\tau_{2,3}$ constrained to the interval $[0,2\,\Delta\rmlabels{t}{}{2,3}]$.
For the dual-feedback oscillator, the times $\delta\tau_{a,r}$ are constrained to the interval $[0,2\,\Delta\rmlabels{t}{}{a,r}]$.
We map values that lie outside these intervals back on it 
by mirroring these values across the nearest boundary of the domain.

Recent single cell experiments revealed the coefficient of variation (CV)
in the time of initiation of DNA replication, $\rmlabels{{\rm CV}}{}{init}=0.7$, 
and in the time of replicating the DNA, $\rmlabels{{\rm CV}}{}{rep}=0.16$,
in slow growing \emph{E. coli} cells \cite{Adiciptaningrum2015}.
Given our models for stochasticity in replication times (including the
fact that the initiation times are constrained to lie in the windows discussed above),
we find that standard deviations of $\rmlabels{\sigma}{}{init}=0.2\td$ and $\rmlabels{\sigma}{}{rep}=0.35\tr$
give similar coefficients of variation.
All parameters are listed in \tref{CHS_Parameters}.


\section{Results}
Here we study how the peak-to-peak times of the
oscillations of the repressilator and the dual-feedback oscillator depend
on the cell division time. 
Furthermore, we illuminate the effects of the position of the
genes on the DNA and the role of stochasticity in the replication
times. Preliminary work on the effect of gene positioning was reported
  in the Supporting Information of \cite{Paijmans2016}.

\subsection{Repressilator}
We first consider the scenario in which the three genes are close
together on the chromosome, such that, to a good approximation, they
are replicated at the same time, and the timing of DNA replication is fixed. 
In \fref{Repressilator_Deterministic}, panel A, we show the
mean peak-to-peak time, \tpeak, in the concentration of $P_1$, for
different cell division times, \td.  Clearly, locking is not very
strong: The locking regions---the range of cell division times where
the mean peak-to-peak time of the repressilator is equal to a multiple
of \td{}---are very small.  The only effect of locking is that in
these very small windows the variance in the peak-to-peak time is
strongly reduced.  The reason why locking is weak is that while the
genes are replicated at the same time, they are expressed at different
times.  This means that gene replication has a different effect on the
expression level of each of the three genes.  Hence, even when the
cell cycle period \td{} is approximately equal to the the oscillator's
intrinsic period \ti, $\td \approx\ti{}$, the oscillation of each
protein concentration has a different amplitude, as shown in
\fref{Repressilator_Deterministic}B.  This makes it harder for all
three protein oscillations to get the same period as that of the cell
cycle, and become locked to it.  Interestingly,
\fref{Repressilator_Deterministic}C shows that when the cell-cycle
time is twice the intrinsic clock period, the pattern of alternating
smaller and larger oscillation amplitudes can still be observed for
each of the respective protein concentration profiles.  This
observation can be used to detect the effect of periodic gene
replication experimentally.

We now consider a scenario in which the different genes are replicated
at different times during the cell cycle, which corresponds to a
situation where the genes are located at different positions on the
chromosome.  We assume that the gene for protein $p_1$ is close to the
origin of replication, such that it is copied at the moment DNA
replication is initiated.  We consider two scenarios for the order of
the genes on the DNA.  In the first scenario, panel D, genes are
placed on the DNA in order of their interaction in the biochemical
reaction network, $p_1, p_2, p_3$ (see \fref{Model_Cartoons}A): The
gene for $p_2$ is copied a time $\Delta t_{1,2}$ after $p_1$, and
$p_3$ a time $\Delta t_{2,3}$ after $p_2$.  In the second scenario,
panel E, genes are in order of maximal expression: $p_3, p_2, p_1$
(see \fref{Repressilator_Deterministic},B and C), which corresponds
to negative values of $\Delta t_{1,2}$ and $\Delta t_{2,3}$.
Throughout this work, we will use the condition $\Delta t_{1,2}=\Delta
t_{2,3}$.  Interestingly, while the locking regions are very small
when the genes are replicated at the same time ($\Delta\rmlabels{t}{}{1,2}=0$, panel A, gray
lines in panels D and E), replicating them at different times
introduces marked locking: both for $\Delta t_{1,2}>0$ (panel D) and
$\Delta t_{1,2}<0$ (panel E) strong locking is observed.  Even more
strikingly, the 1:1 locking region is largest when genes are
replicated in order of maximal expression, and when the distance
between them is the largest (panel E).  This can be understood by
noting that when genes are replicated in the order of maximal
expression, shifting the phase of the clock with respect to that of
the cell cycle has then the strongest effect on the amplitude and
hence the period of the clock oscillations, which underlies the
phenomenon of locking, as explained in \cite{Paijmans2016}.

\begin{figure}[ht!]
\center
\includegraphics[scale=1.0]{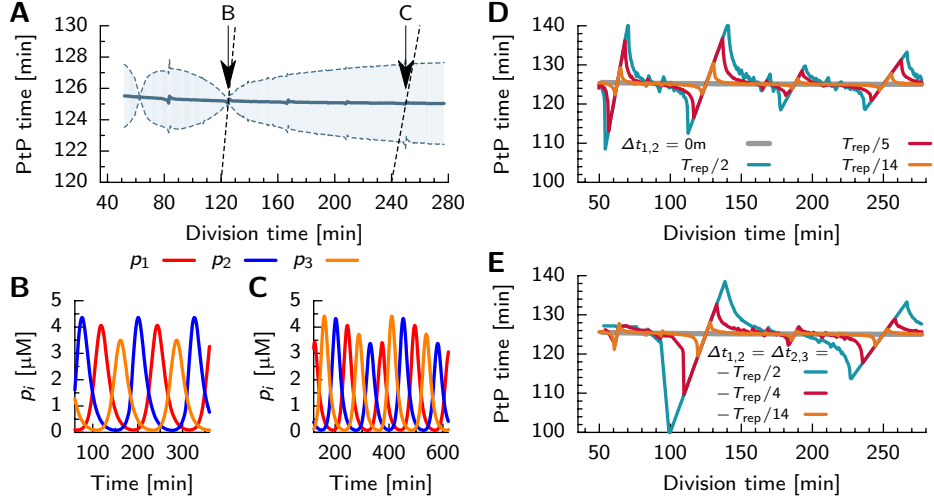}
\caption{\flabel{Repressilator_Deterministic} The repressilator \cite{Elowitz2000} 
  can strongly lock to the cell cycle, and the strength of locking depends sensitively on how the genes are positioned on the DNA.
  (A) Average (solid line) and standard deviation (shaded region) of
  the peak-to-peak time \tpeak\ as a function of the division time,
  where the time between replicating genes is $\Delta t_{1,2}=\Delta t_{2,3}=0$.
  The repressilator has an intrinsic period of $T_{\mathrm{int}} = 125$.  
  The locking regions around $T_{\mathrm{int}}$ and $2T_{\mathrm{int}}$
  are almost absent.  (B and C) Representative time traces of the
  concentrations of the three repressilator proteins, $p_1(t)$ (red),
  $p_2(t)$ (blue) and $p_3(t)$ (orange), for the cell-division times
  indicated by the arrows in panel A.  
  (B) When $\td=T_{\mathrm{int}}$, the oscillations are very regular (almost no
  variance in the PtP-times), but each protein concentration has a different amplitude. 
  (C) At $\td=2T_{\mathrm{int}}$, all three protein concentrations switch between a small and a
  large amplitude in successive oscillation cycles.  
  Panels D ($\Delta t_{1,2}>0$) and E ($\Delta t_{1,2}<0$) show the effect of varying the timing of replication of the three genes,
  assuming $\Delta t_{2,3}=\Delta t_{1,2}$.
  For clarity, we only show the average peak-to-peak time as a function of \td, 
  not the standard deviation. 
  Values of $\Delta t_{1,2}$ are given in the legend, 
  and are written as a fraction of the mean DNA replication time \tr.
  Remarkably, for all $\Delta t_{1,2}\neq 0$, there is significant locking.  
  Clearly, the timing of gene replication can markedly affect locking, 
  which means that the spatial distribution of the genes over the chromosome can
  be of critical importance in the interaction between the clock and the cell cycle. 
  (Figure adapted from \cite{Paijmans2016})}
\end{figure}

To see if locking persists in the presence of physiologically levels of noise in gene replication times,
we change the gene replication times $t_1, t_2$ and $t_3$ into stochastic variables via \eref{RepTimesStochComb}.
Our results reveal that both when $\Delta t_{1,2}>0$ (\fref{Repressilator_StochasticCombined}, panel A) 
and when $\Delta t_{1,2}<0$ (panel B),
the coupling of the repressilator to the cell cycle is strongly attenuated.
However, the effects of the cell cycle are still clearly observable 
around the 1:1 locking region and when $\td=0.5\ti$.
For division times longer than the oscillators intrinsic period, 
all signatures of locking have disappeared.

\begin{figure}[ht!]
\center
\includegraphics[scale=1.0]{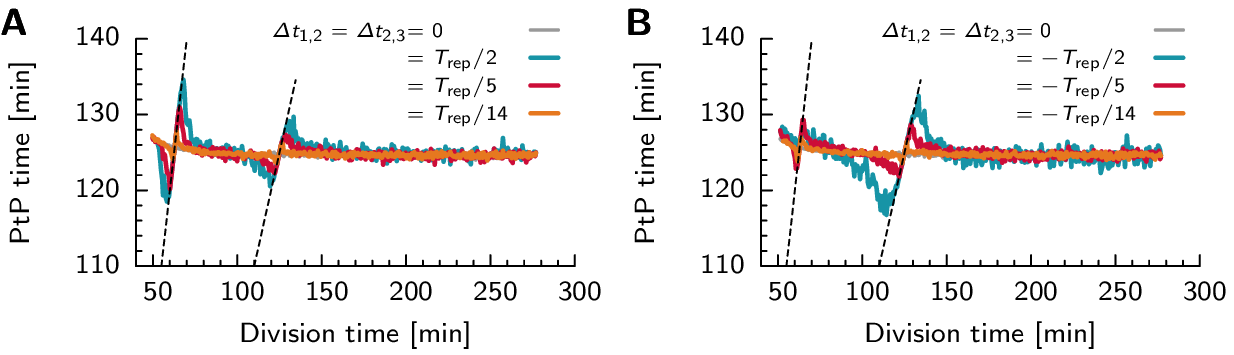}
\caption{\flabel{Repressilator_StochasticCombined} In the repressilator, 
locking persists in the presence of physiological levels of noise in the gene replication times.
In all panels, solid lines show the peak-to-peak time in the activator
concentration for different periods of the cell division time \td. 
Standard deviation in \tpeak{} omitted for clarity, but is similar in all panels.
Legends are defined in \fref{Repressilator_Deterministic}.
We used the physiologically motivated values for the standard deviations
in the timing of the initiation, $\rmlabels{\sigma}{}{init}=0.2\td$,
and the progression, $\rmlabels{\sigma}{}{rep}=0.35\tr$, of DNA replication.
The two panels show a different order of the genes $p_1,p_2$ and $p_3$ 
with $\Delta t_{1,2}>0$ (A), and $\Delta t_{1,2}<0$ (B).
Clearly, at these noise levels,
locking is strongly reduced compared to the deterministic case (See \fref{Repressilator_Deterministic}, panels D and E),
but still clearly observable around $\td=\ti$.
}
\end{figure}

\subsection{Dual-feedback oscillator}
\fref{Hasty_Deterministic}A shows strong locking of the dual-feedback
oscillator to the cell cycle.  We assume here that the genes are
located next to each other on the chromosome, so that their
time-varying gene-densities are the same.  Clearly, the widths of the
locking regions are very large; they are even larger than those
observed for our simple negative feedback oscillator studied in \cite{Paijmans2016}. 
In \fref{Hasty_Deterministic}B we show a time trace
of the irregular oscillations around a cell-division time of $\td = 98$ minutes. 
\fref{Hasty_Deterministic}C shows that the amplitude of the
oscillations alternates between a high and a low value when the
cell-division time \td{} is about twice the intrinsic clock period of
$\ti=74$ minutes, due to periodic gene replication every other clock period. 
We thus conclude that also the dual-feedback oscillator can
strongly lock to the cell cycle and that this effect should be
observable experimentally.

\fref{Hasty_Deterministic}D,E shows the result of varying the moment of gene replication for the two genes. 
Again, in this model, the first gene of the oscillator is placed next to the origin of replication 
such that it is replicated at initiation of DNA replication,
and the second gene is replicated with a mean delay $\Delta\rmlabels{t}{}{a,r}$ later.
For positive $\Delta\rmlabels{t}{}{a,r}$, the activator is replicated before the repressor,
and negative $\Delta\rmlabels{t}{}{a,r}$, vice versa.
We vary the time delay between the replication of the two genes, 
as $\Delta\rmlabels{t}{}{a,r}=0,\tr/8,\tr/2$ and \tr{} (panel D) 
and minus these values (panel E), where \tr{} is the mean replication time of the DNA.
It is seen that in both scenarios the strength of locking 
decreases with increasing the distance between the genes on the DNA:
The strongest entrainment is observed when the genes are replicated at the same time during the cell cycle (gray lines),
in stark contrast to the behavior of the repressilator.
While in the repressilator the locking increases with the distance between the genes,
the dual-feedback oscillator shows the opposite behavior. 
Interestingly, though, in the dual-feedback oscillator locking still persists
when the genes are placed at maximum distance from each other.

\begin{figure}[ht!]
\center
\includegraphics[scale=1.0]{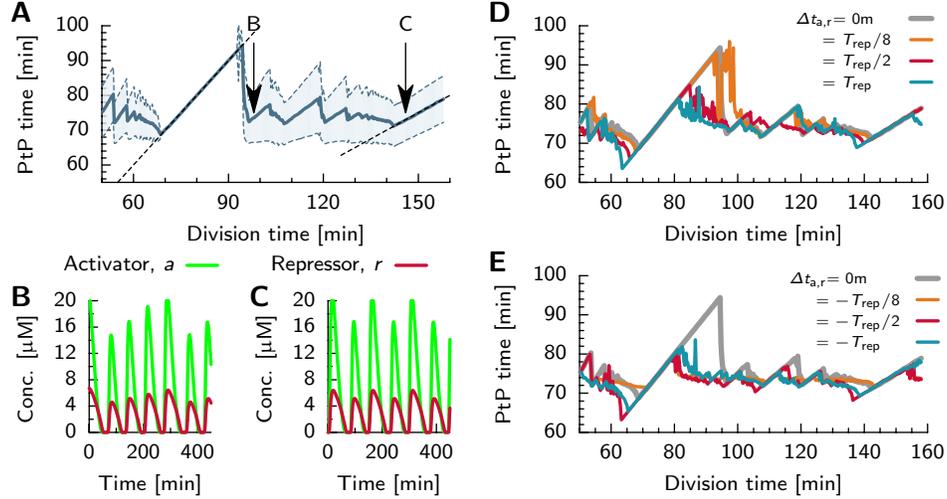}
\caption{\flabel{Hasty_Deterministic} The dual-feedback oscillator
  \cite{Stricker2008} can strongly lock to the cell cycle, and the
  strength of locking depends on the temporal order in which the
  genes are replicated during the cell cycle.  
  The intrinsic period of the oscillator $T_{\mathrm{int}}=73$ minutes.  
  (A) Average (solid line) and standard deviation (shaded region) of the peak-to-peak
  time \tpeak\ as a function of the division time \td\,
  when both genes are replicated simultaneously, $\Delta t_{\mathrm{a,r}}=0$. 
  There is a wide region of cell division times (around $\td=\ti$) 
  where the oscillator has a \tpeak\ equal to the the cell cycle (left dashed line). 
  (B and C) Representative time traces for the division times indicated by the arrows in panel A. 
  Shown are the activator and repressor concentrations $a(t)$ (green line) and $r(t)$ (green line), respectively. 
  At a cell-division time of $\td=98$ min (B), 
  just outside the region where the oscillator is locked to the cell cycle,
  the time traces show very irregular behavior resulting in a large variance in the PtP times. 
  At $\td=2T_{\mathrm{int}}$ (C), 
  the oscillations switch between a small and a large amplitude in successive oscillation cycles, 
  a signature of the periodic gene replications.  
  (D and E) The effect of the order of gene replication during the cell cycle. 
  For clarity, only the average peak-to-peak time as a function of \td\ is shown,
  not the standard deviation. 
  Values of $\Delta t_{\mathrm{a,r}}$ are given in the legend, 
  and are written as as a fraction of the DNA replication time \tr{}.  
  (D) Positive $\Delta t_{\mathrm{a,r}}$; the repressor gene is replicated after the activator gene.
  (E) Negative $\Delta t_{\mathrm{a,r}}$; the repressor gene is replicated before the activator gene. 
  Remarkably, contrary to the behavior of the Repressilator,
  locking decreases with increasing time delay between replicating genes $\Delta t_{\mathrm{a,r}}$.
  This illustrates that the influence of the cell cycle on the clock depends in a non-trivial way on the
  architecture of the clock and on the nature of the driving signal. (Figure adapted from \cite{Paijmans2016})}
\end{figure}

To see if locking persists in the presence of noise in the timing of gene replications, 
we changed the time of replication of both genes, \rmlabels{t}{}{a} and \rmlabels{t}{}{r}, 
into stochastic variables via \eref{RepTimesStochComb}. 
The noise strongly attenuates the effects of the cell cycle, 
both for positive (\fref{Hasty_StochasticCombined} panel A) and negative (panel B) $\Delta\rmlabels{t}{}{a,r}$, 
as compared against deterministic result of \fref{Hasty_Deterministic}. 
However, the peak-to-peak times of the dual-feedback oscillator are still
perturbed around the 1:1 locking region, especially in the case $\Delta\rmlabels{t}{}{a,r}>0$.

\begin{figure}[ht!]
\center
\includegraphics[scale=1.0]{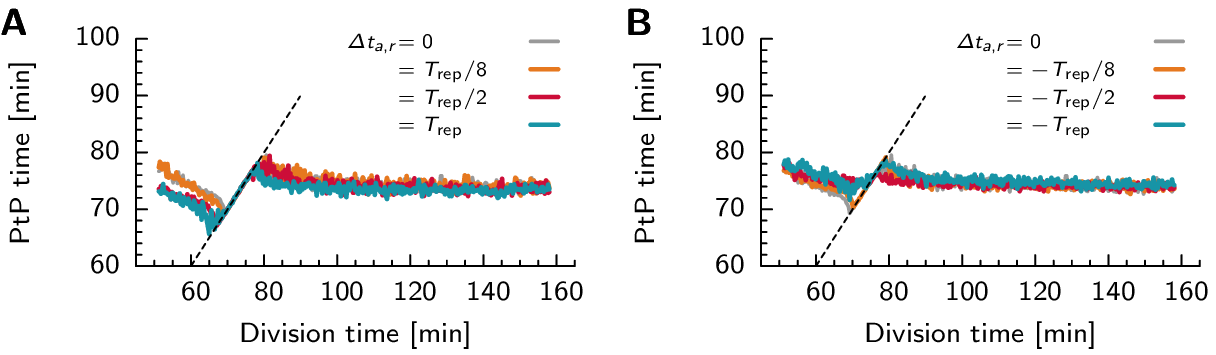}
\caption{\flabel{Hasty_StochasticCombined} In the dual-feedback oscillator, 
locking persists in the presence of physiological levels of noise in the gene replication times.
In all panels, solid lines show the peak-to-peak time in the activator
concentration for different periods of the cell division time \td. 
Standard deviation in \tpeak{} omitted for clarity, but is similar in all panels.
Legends are defined in \fref{Hasty_Deterministic}.
We used the physiologically motivated values for the standard deviations
in the timing of the initiation, $\rmlabels{\sigma}{}{init}=0.2\td$,
and the progression, $\rmlabels{\sigma}{}{rep}=0.35\tr$, of DNA replication.
The two panels show a different order of the activator and repressor 
gene with $\Delta t_{\mathrm{a,r}}>0$ (A), and $\Delta t_{\mathrm{a,r}}<0$ (B).
As observed for the repressilator, 
locking is strongly reduced compared to the deterministic case (See \fref{Hasty_Deterministic}, panels D and E),
but still clearly observable around $\td=\ti$.}
\end{figure}

\subsection{What attenuates the effects of the cell cycle more:
Stochasticity in the initiation or progression of DNA replication?}
Comparing \fref{Repressilator_Deterministic} with \fref{Repressilator_StochasticCombined} for
the repressilator and \fref{Hasty_Deterministic} with \fref{Hasty_StochasticCombined} for 
the dual-feedback oscillator,
it is clear that noise in gene replication times has a significant effect on the coupling
between the cell cycle and these synthetic oscillators.
In our model, noise in the replication times is the result of noise in the initiation 
and in the progression of DNA replication.
We want to know which of these two sources of stochasticity is key for reducing the coupling 
between the cell cycle and the oscillator.

To find out whether the initiation or the progression of DNA replication is more important
for attenuating the effects of gene replications,
we studied two models for the noise in the replication times.
Either there is only noise in the progression of replication, 
such that the time intervals between replicating different genes, $\delta\rmlabels{\tau}{}{1,2}$ and $\delta\rmlabels{\tau}{}{a,r}$,
are stochastic variables but the time of initiation of DNA replication is deterministic, 
$\Delta\rmlabels{t}{}{init}=\rmlabels{\alpha}{}{init}\td$ (See \fref{CellCycle_Cartoons}C).
Or, the initiation of DNA replication, $\delta\rmlabels{\tau}{}{inti}$, is stochastic but the 
progression of replication is deterministic such that $\Delta\rmlabels{t}{}{1,2}$ and $\Delta\rmlabels{t}{}{a,r}$ 
are fixed each cell cycle (See \fref{CellCycle_Cartoons}D).
We will use the same values for the standard deviations $\rmlabels{\sigma}{}{init}$ and $\rmlabels{\sigma}{}{rep}$ 
of the two noise sources as before.

In \fref{Repressilator_StochasticSeperate} we show the effects of the cell cycle on the period of the repressilator
when there is only noise in the progression of replication, panels A and B,
or when there is only noise in the initiation of DNA replication, panels C and D.
Clearly, when there is only noise from replication progression, 
both for positive (panel A) and negative (panel B) $\delta\rmlabels{\tau}{}{1,2}$,
the width of the locking regions are almost the same as compared to 
the deterministic case (See \fref{Repressilator_Deterministic}, panels D and E).
The effects of the cell cycle are not significantly attenuated by the noise in DNA replication progression.
However, when the noise is due to the initiation of replication 
(panels C and D for positive and negative $\delta\rmlabels{\tau}{}{1,2}$, respectively),
all signatures of coupling disappear for $\td>\ti$, and the width of the 1:1 locking region 
is strongly reduced compared to the case of a deterministic cell cycle.
We conclude that the decrease in locking to the cell cycle is predominantly 
due to the stochasticity in the initiation time of DNA replication.

For the dual-feedback oscillator we obtain similar results. 
In \fref{Hasty_StochasticSeperate} we show the effects of the cell cycle on the period of the dual-feedback oscillator
when there is only noise in the progression of replication, panels A and B,
or when there is only noise in the initiation of DNA replication, panels C and D.
When there is only noise due to the progression of DNA replication,
both for positive (panel A) and negative (panel B) $\delta\rmlabels{\tau}{}{a,r}$,
strong signatures of locking persists, especially around $\td=\ti$ and $\td=2\ti$.
Again, stochasticity in DNA replication progression does not attenuate the coupling to the cell cycle much.
When the source of noise is due to stochasticity in the initiation of DNA replication,
almost all effects of the cell cycle on the peak-to-peak time of the dual-feedback oscillator 
have disappeared; only when $\td=\ti$ locking can still be observed.
Clearly, also for the dual-feedback oscillator the initiation of DNA replication has the biggest effect on the
coupling between the cell cycle and the oscillator.

We observe that, both for the repressilator and the dual-feedback oscillator, 
the initiation of DNA replication is dominant in attenuating the effects of the cell cycle.
Why is this the case?
An oscillator couples to the cell cycle by maintaining a specific phase relation between
the phase of the oscillator and that of the gene density, as explained in \cite{Paijmans2016}.
When the standard deviation in the replication times is of the same order as the intrinsic
period of the oscillator, it becomes impossible to maintain this phase relation,
and the oscillator can not couple to the cell cycle.
Because in our model, the standard deviation in the initiation of replication is proportional to \td{},
while the standard deviation in replication \emph{progression} is constant, 
initiation of DNA replication will be the dominant source of noise when $\td>\ti$.
Indeed, for $\td>\ti$, the stochasticity in the initiation of DNA replication will be so large, 
that the clock no longer couples to the cell cycle (See \fref{Repressilator_StochasticCombined} and \fref{Hasty_StochasticCombined}). 
For $\td\leq\ti$, the stochasticity in the initiation of DNA replication is much smaller. 
Moreover, the noise in DNA replication progression is so small that the coupling of the clock to the cell cycle 
is not much weakened by it (See \fref{Repressilator_StochasticSeperate}A,B and \fref{Hasty_StochasticSeperate}A,B). 
This explains why for $\td\leq\ti$, noise in DNA replication does not appreciably attenuate the locking of the clock 
to the cell cycle.

\begin{figure}[t!]
\center
\includegraphics[scale=1.0]{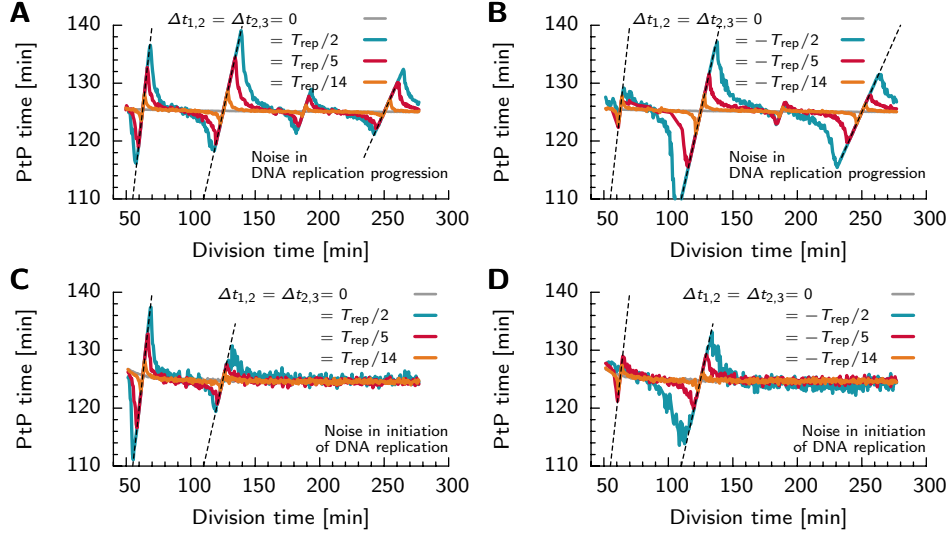}
\caption{\flabel{Repressilator_StochasticSeperate} In the repressilator,
  stochasticity in the initiation of DNA replication plays the dominant role in attenuating 
  the effects of gene replications.
  Top row, panels A and B, show results with only noise in the progression of DNA replication, 
  $\rmlabels{\sigma}{}{rep}=0.35\tr$,
  and the bottom row, panels C and D, corresponds to the situation where there is only noise in replication initiation, 
  $\rmlabels{\sigma}{}{init}=0.2\td$ (see \fref{CellCycle_Cartoons}). 
  In both panels, solid lines show the peak-to-peak time \tpeak{} 
  in the oscillations of $P_1$ as a function of the cell division time \td. 
  Standard deviation in \tpeak{} omitted for clarity, 
  but is similar in all panels. 
  Legends are defined in \fref{Repressilator_Deterministic}. 
  (A,B) When there is noise in the time intervals between
  the gene replication events, but the initiation of DNA replication
  is fixed, locking seems little affected compared to the
  deterministic case (See \fref{Repressilator_Deterministic}, panels D and E).
  (C,D) When there is noise in the initiation of DNA replication, 
  but the time between replications is fixed,
  the effects of the cell cycle almost disappear for division times $\td>\ti$, 
  in both ways of ordering the genes. 
  However, strong locking persists at the 1:1 locking region
  and for $\td<\ti$.
  Comparing with panels A and B, noise in the initiation of
  DNA replication seems to be more effective in protecting the clock
  against the cell cycle.}
\end{figure}

\begin{figure}[ht!]
\center
\includegraphics[scale=1.0]{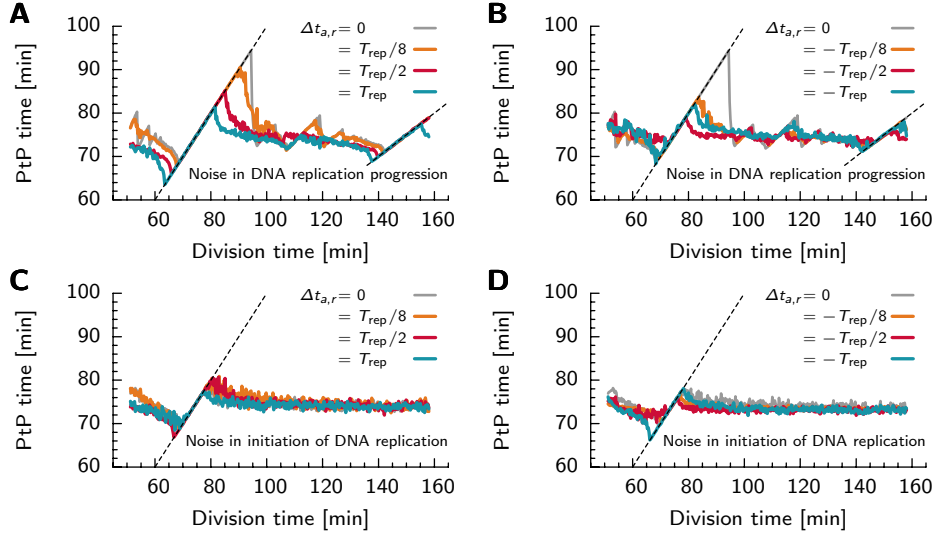}
\caption{\flabel{Hasty_StochasticSeperate} In the dual-feedback oscillator,
  stochasticity in the initiation of DNA replication plays the dominant role in attenuating 
  the effects of gene replications.
In both panels, solid lines show the peak-to-peak time in the activator
concentration for different periods of the cell
division time \td. 
Standard deviation in \tpeak{} omitted for clarity, but is similar in all panels.
Legends are defined in \fref{Hasty_Deterministic}.
We compare a scenario with only noise in DNA replication progression, 
with a standard deviation $\rmlabels{\sigma}{}{rep}=0.35\tr$, panels A and B,
to a scenario with only noise in the initiation of DNA replication,
with a standard deviation $\rmlabels{\sigma}{}{init}=0.2\td$, panels C and D.
(A,B) As observed for the repressilator, with noise in replication
progression but not in replication initiation,  
locking is little affected, compared to the deterministic case (See \fref{Hasty_Deterministic}, panels D and E).
(C,D) In the opposite scenario, with noise in the initiation of DNA replication
but not in the progression of replication,
most signatures of locking disappear, 
both when the activator or repressor gene is replicated first.
Only around $\td{}=\ti$, locking persists.
Clearly, comparing with panels A and B,
noise in the initiation of DNA replication has a stronger attenuating effect on locking.
}
\end{figure}

%
%
%
%

\section{Discussion}
Discrete gene replication events, present in all cells, 
can have marked effects on the period of circadian clocks \cite{Paijmans2016}.
We wanted to know how gene replications affect the robustness 
of two renowned synthetic oscillators build in \emph{E. coli}:
The repressilator by Elowitz \textit{et. al.} \cite{Elowitz2000} and
the dual feedback oscillator by Stricker \textit{et. al} \cite{Stricker2008}.
Using computational modeling, we show how the peak-to-peak time of the oscillators depend on the
cell division time, the position of the genes on the DNA and the noise in the gene replication times.

We find that both synthetic oscillators can strongly lock to the cell
cycle, where the oscillator's peak-to-peak time is equal to a multiple
of the cell division time, over a wide range of division times.
Remarkably, the effect strongly depends on how the genes of the
oscillator are located on the chromosome.  The distance between the
genes introduces a temporal delay between the moments at which the
different genes of the oscillators are replicated, which affects the
period of the oscillations.  Increasing the distance between genes has
an opposite effect on the two oscillators: Whereas the repressilator
exhibits almost no locking when the genes are positioned close
together yet strong coupling over a wide range of \td{} when
the temporal delay is increased, the dual-feedback oscillator shows the
strongest coupling to the cell cycle at negligible temporal delay
between gene replications. 
For both models, the signature of the gene
replication events should be clearly visible in the amplitude of the
time traces of the protein concentrations.

It is well known that the timing of key events during the cell cycle,
such as the start of DNA replication, the duration of chromosome
replication and cell division, exhibit high levels of stochasticity
\cite{Koppes1980, Michelsen2003}, which will propagate to the
replication times of the oscillator's genes.  To investigate how
strong noise in the timing of gene replication affects the
oscillator's coupling to the cell cycle, we introduced two noise sources 
in our model of the cell cycle: one in the time of when DNA replication is initiated
and one in the time it takes to replicate the chromosome.
Using physiologically relevant values for the standard
deviations in the timing, we found that noise in gene replication times
strongly attenuates the effects of the cell cycle.
However, observable signatures of locking remain for division times
equal and shorter than the oscillator's intrinsic period.  For these
cells, the standard deviation in gene replication times becomes
smaller than the oscillator's intrinsic period, making it possible for
the clock to lock to a certain phase of the gene density, which sets
the peak-to-peak time. 
We then asked which of these two sources is more important in attenuating the coupling 
between the cell cycle and the oscillator.
To this end, we made two models for stochasticity in the replication times:
One with only noise in replication progression and the other with only 
noise in the time of replication initiation.
We found that noise in the initiation of DNA replication has a stronger effect 
than that in the progression of DNA replication.  
The reason is that, at physiologically motivated values,
the standard deviation in the time of replication initiation is much larger
than the standard deviation in the time of replicating the chromosome.
We thus conclude that the initiation of DNA replication is mainly
responsible for attenuating the effects of the gene replications on the 
repressilator and dual-feedback oscillator.


Throughout this work, we assume the genes reside on the bacterial chromosome. 
Importantly, however, the synthetic oscillators were
originally constructed on plasmids, which are often present in large copy
numbers ranging from 10-100.  Moreover, experiments indicate that these
plasmids are copied at random times during the major part of the cell
cycle \cite{Bogan2001}. Based on our observation that multiple
chromosome copies that are replicated asynchronously strongly reduce
the strength of locking \cite{Paijmans2016},
we expect that, at these high
plasmid copy numbers, the synthetic oscillators exhibit no clear
signatures of locking. Indeed,  the original study on the
dual-feedback oscillator does not report any effects from the cell
cycle, even when the growth rate is comparable to the oscillator's
intrinsic period where locking is expected to occur
\cite{Stricker2008}. 
Remarkably, however, signatures of locking were observed for the
dual-feedback oscillator in the experiments of \cite{Dies2015},
even though also in these experiments the genes reside on plasmids \cite{Dies2015}.
It is hard to explain what underlies the effect of the cell cycle in
these experiments \cite{Dies2015}. In any
case, our analysis predicts that  the effects will be much stronger when the genes are put on the chromosome.
Conversely, in order to prevent locking, 
it seems beneficial to construct the oscillator on high copy number plasmids.

The genes of biological oscillators such as circadian clocks do reside
on the chromosome, and the periods of these oscillators are often
unaffected by the cell cycle \cite{Mori2001}.  One approach to
  understand how these natural clocks are so resilient to
  perturbations from the cell cycle is to construct synthetic
  oscillators in growing and dividing cells.  
The dual-feedback oscillator studied in this work, 
based on a coupled positive and negative feedback architecture regulating gene expression, 
has been predicted to produce robust oscillations \cite{Tsai2008,Woods2016}:
The amplitude and period do not critically depend on specific parameter values, 
and oscillations persist in a wide range of temperatures and growth media
\cite{Stricker2008,Prindle2014}.  
However, these models do not take the effect of gene replications into account,
and in the experiments the genes reside on high copy-number plasmids,
potentially abolishing any effect of the cell cycle.  Our results suggest that the
relatively simple design of the dual-feedback oscillator implemented
on the chromosome might not be very robust in growing and dividing
cells, since its period scales with that of the cell
cycle. Clearly, to test the predictions of our analysis, it would
  be of interest to implement this oscillator on the chromosome,
  which is now increasingly being done in synthetic biology \cite{Sowa2015}.
Comparing the unstable synthetic oscillators with their
evolved stable counterparts found in, e.g. \emph{S. elongatus} and
\emph{N. crassa}, could elucidate why the latter feature a remarkably
more complex reaction network, including, for example,
post-translational modification of the proteins
\cite{Paijmans2016,Hurley2016}.

\section{Methods}
Both the models of the repressilator and dual-feedback oscillator are described with ordinary differential equations,
and propagated using Mathematica 8 (Wolfram Research).  
For each value of $T_\text{d}$, we generate a single time trace of about 200 oscillations 
for the repressilator and 100 oscillations for the dual-feedback oscillator.
In order to allow the oscillations to settle down to a steady state, 
we discard the first 10 oscillations in the system.

To simulate the (stochastic) gene replication events, for each gene $n$ in the model, 
we generate a list of replication times, $\tau_i^n$, using \erefs{RepTimesDet}-\ref{eqn:RepTimesStochComb}.
The gene copy number for this gene, $g^n(t)$, equals 1 when $t<\tau_i^n$,
and 2 when $t>\tau_i^n$, modulo \td.
The discrete gene replication events enter the models via the gene density, $G^n(t)=g^n(t)/V(t)$, 
where $V(t)=\mathrm{exp}(\mathrm{ln}(2)/\td\,\mathrm{mod}(t,\td))$ is the cell volume \cite{Paijmans2016}.

To find the peak-to-peak times, \tpeak, in the ODE simulations 
(including those with noise in the gene replication times), 
we use the built-in methods of Mathematica to return all local
extrema in the concentration of $p_1$ (repressilator) or the activator (dual-feedback oscillator).
These extrema correspond to the time points $t_i$ where the concentration 
is higher, in the case of a maximum, or lower, in the case of a minimum, than its two immediate neighbors.  
As is standard for numerical solution of differential equations, 
the spacing $t_{i}-t_{i-1}$ between successive time points is determined 
adaptively by the algorithm to meet imposed precision bounds but never exceeded 0.2 h.
We then checked if a given local minimum is the lowest point 
within an interval of $\pm$ 3/4 the oscillator's intrinsic period, \ti, centered on the minimum; 
if so, we define this point as the global minimum of a single oscillation cycle.  
If there exist a local extremum with a lower value,
we repeated this procedure around the lower point until we found a point which 
was the lowest within a time interval of $\pm$ $3/4\ti$.
The same procedure is used for finding the local maxima of the oscillations. 
The peak-to-peak time is then calculated by subtracting the times of two consecutive minima; 
we verified that subtracting the times of two consecutive {\em maxima} gave essentially the same results.

\begin{acknowledgments}
  We thank Giulia Malaguti for a critical reading of the manuscript. 
  This work is part of the research programme of the Netherlands Organisation for Scientific Research (NWO).
  (JP and PRtW), and by NSF Grant DMR-1056456 (DKL).
\end{acknowledgments}


\end{document}